\documentclass[12pt]{article}
\usepackage{amssymb,amsmath,amsthm,amsfonts,amscd}
\usepackage{graphicx}
\newcommand\be{\begin{equation}}
\newcommand\ee{\end{equation}}

\newcommand{\fatalpha}{{\bf \alpha \kern -0.44em \alpha}}
\newcommand{\fatsigma}{{\bf \sigma \kern -0.54em \sigma}}
\newcommand{\tpchi}{{\bf \chi \kern -0.35em \chi}}
\newcommand{\llambda}{{\bf \lambda \kern -0.45em \lambda}}

\bibliography{plain}
 \title{\bf  Noise Effects on Entangled Coherent State Generated via Atom-Field Interaction and Beam Splitter}
 \vspace{20mm}
\author{G. Najarbashi \thanks{E-mail: Najarbashi@uma.ac.ir} ,
  S. Mirzaei \thanks{E-mail: SMirzaei@uma.ac.ir}
\\
{\small Department of Physics, University of Mohaghegh Ardabili, P.O. Box 179, Ardabil, Iran.} \\
 \\
}\pagebreak
\begin{document}
\maketitle \vspace{0mm}

\maketitle \vspace{0mm}
\begin{abstract}
In this paper, we introduce a controllable method for producing two and three-mode entangled coherent states (ECS's) using atom-field interaction in cavity QED and beam splitter. The generated states play central roles in linear optics, quantum computation and teleportation. We especially focus on qubit, qutrit and qufit like ECS's and investigate their entanglement by concurrence measure. Moreover, we illustrate decoherence properties of ECS's due to  noisy channels, using negativity measure. At the end the effect of noise on monogamy inequality is discussed.
\end{abstract}
\newpage
\section{Introduction}
 Coherent states originally introduced by Schrodinger in 1926 \cite{Schrodinger} and then advanced studies were done in \cite{Glauber,Perelomov}. They have played an important role in different problems in quantum information theory. In recent years, much attention has been devoted to the problem of generating various quantum states of an electromagnetic field \cite{Enk2,Sanders3,Milburn1,Milburn2,Tanas1,Tanas2,Tanas3,gerry}. In \cite{Zeng}, an alternative scheme to produce superpositions of a series of coherent states on a circle in the phase space with only one atom driven by a classical field is presented. A new scheme for preparation of a type of nonclassical state in cavity QED, which we use in this work, is proposed in \cite{zhen}. In this scheme, an atom either flying through or trapped within a cavity, is controlled by the classical Stark effect.\\
 Generation of multipartite ECS's and entanglement of multipartite states constructed by linearly independent coherent states are investigated in \cite{Barry,Enk1,najarbashi1}. On the other hand ECS's have many applications in quantum optics and quantum information processing \cite{Cochrane,Oliveira,Kim,Milburn,Munro,wang3,wang5,wang1,Vogel1,Salimi}.  van Enk and Hirota \cite{Enk} discussed how to teleport a Schrodinger cat state through a quantum channel described by the maximally ECS. In \cite{wang2,wang4} the required conditions for the maximal entangled states of the form $|\psi\rangle=\mu|\alpha\rangle|\beta\rangle+\nu|\gamma\rangle|\delta\rangle$ have been studied and this subject have been generalized to the state $ |\psi\rangle=\mu|\alpha\rangle|\beta\rangle+\lambda|\alpha\rangle|\delta\rangle+
 \rho|\gamma\rangle|\beta\rangle+\nu|\gamma\rangle|\delta\rangle$  in Ref. \cite{najarbashi}.\\
Another problem which has been investigated extensively in quantum information processing is noise effect or decoherence which arise from the coupling of the system to its surroundings \cite{Enk1,Enk,Yao}. van Enk in \cite{Enk} introduced
the effect of noise on coherent states with the modes $1$ or $2$ after traveling through a noisy channel as
\be
|\alpha\rangle_{1(2)}|0\rangle_E\rightarrow|\sqrt{\eta}\alpha\rangle_{1(2)}|\sqrt{1-\eta}\alpha\rangle_E
\ee
where the second state now refers to the environment and $\eta$ is the noise parameter, which
gives the fraction of photons that survives the noisy channel. Moreover at the same work, the noise effect on teleportation fidelity was established. Dynamics of maximum ECS and optical qubits which is encoded via coherent states subject to environmental noise was studied in \cite{Wu,El}.\\
In this paper, following Ref. \cite{zhen} we use a controllable method for generating the superposition of two, three and four glauber states using Jaynes-Cummings model. Then we propose a new scheme for producing two and three-mode ECS's. Assuming the linearly independent of coherent states, we introduce qubit, qutrit and qufit like ECS's. For two-mode qubit like ECS as
\be
|\Psi^{(2)}\rangle
=\frac{1}{\sqrt{M^{(2)}}}(|\frac{-\alpha}{\sqrt{2}}\rangle_{_{1}}|\frac{-\alpha}{\sqrt{2}}\rangle_{_{2}}-|\frac{\alpha}{\sqrt{2}}\rangle_{_{1}}|\frac{\alpha}{\sqrt{2}}\rangle_{_{2}}),
\ee
if the set $\{|\alpha\rangle,|-\alpha\rangle\}$ are linearly independent meaning they span a two dimensional Hilbert space $\{|0\rangle,|1\rangle\}$, the two-mode coherent state $|\Psi^{(2)}\rangle$ can be recast in two qubit form. The same argument is hold for qutrit like coherent states $|\Psi^{(3)}\rangle$. By considering that three coherent states are in general nonorthogonal, i.e. they span a three dimensional Hilbert space $\{|0\rangle,|1\rangle,|2\rangle\}$, the state $|\Psi^{(3)}\rangle$ recast in two qutrit form. The same processing is used for qufit like ECS's $|\Psi^{(4)}\rangle$. The entanglement between modes $1$ and $2$ can be calculated using concurrence measure. Moreover we investigate noise effect on entanglement between modes $1$ and $2$ by negativity and show that it is decreased after traveling through the noisy channel. Finally, we will discuss  the effect of noise on the monogamy inequality for three-qubit entangled states.\\
The outline of this paper is as follows: In section 2 we propose a scheme for the generation of qubit, qutrit and qufit like ECS's using atom-field interaction in cavity QED and beam splitter. A suitable measure which is used for quantifying the entanglement is concurrence. In section 3 we investigate decoherence properties of ECS due to lossy channels, using negativity measure. The effect of noise on monogamy inequality is probed is section 4. The main conclusions of this paper are presented in Section 5.
\section{Generation of ECS's}
In this paper, we use cavity QED for generating superpositions of coherent states and propose a new scheme for preparation of ECS's via beam splitter.
\subsection{Atom-Field Interaction in Cavity QED for Generating Superpositions of Coherent States}
To begin, let us suppose that the hamiltonian of simplified model of the matter-light interaction as
\be
\hat{H}_{int}=\omega_{c}\hat{a}^\dag \hat{a}+\omega_{a}\hat{\sigma}_{z}+g(\hat{a}^\dag \hat{\sigma}^{-}+\hat{a}\hat{\sigma}^{+})+\varepsilon e^{-i\omega_{L}t}\hat{\sigma}^{+}+\varepsilon^{*} e^{i\omega_{L}t}\hat{\sigma}^{-},
\ee
i.e. the interaction of a two-level atom with a single mode of the radiation field, which is known as the Jaynes-Cummings model \cite{Zeng,zhen}, where $\hat{a}(\hat{a}^{\dag})$ is the annihilation (creation) operator for the cavity field and for simplicity we assume that $\hbar=1$. $\hat{\sigma}^{\pm}$ and $\hat{\sigma}_{z}$ are the atomic transition operators which are given by
\be
\begin{array}{l}
\hat{\sigma}^{+}=|e\rangle\langle g|,
\hat{\sigma}^{-}=|g\rangle\langle e|,\\
\hat{\sigma}_{z}=|e\rangle\langle e|-|g\rangle\langle g|,
\end{array}
\ee
and $\omega_c$, $\omega_L$ and $\omega_a$ are the frequencies of the cavity, classical field and the atomic transition frequency between the excited state $|e\rangle$ and the ground state $|g\rangle$ respectively. $g$ is the atom-cavity
coupling constant. The complex amplitude is represented by $\varepsilon$. We assume that the atom is not affected by the cavity field and is initially resonant with the classical field. The interaction hamiltonian for such system is written as
\be\label{1}
\hat{H}_{I1}=\varepsilon e^{-i\varphi}\hat{\sigma}^{+}+\varepsilon^{*} e^{i\varphi}\hat{\sigma}^{-}.
\ee
If the atom is initially in the ground state $|g\rangle$, the hamiltonian of Eq.(\ref{1}) leads to the following transition:
\be
\begin{array}{l}
|g\rangle \rightarrow \frac{1}{\sqrt{1+|\varepsilon_{k}|^2}}(|g\rangle+\varepsilon_{k}|e\rangle),\\
|e\rangle\rightarrow \frac{1}{\sqrt{1+|\varepsilon_{k}|^2}}(-\varepsilon_{k}^{*}|g\rangle+|e\rangle),
\end{array}
\ee
where $\varepsilon_{k} (k=0,1,2,...)$ is an adjustable complex number controlled by the parameters of the classical field. Now if the atom is interacting dispersively with the cavity field and far away from the classical field, the effective hamiltonian of the atom-cavity system is given by
\be\label{2}
\hat{H}_{I2}=\frac{g^2}{\Delta}\hat{a}^{\dag} \hat{a}\hat{\sigma}_z,
\ee
in which $\Delta=\omega_a-\omega_c$. We assume that the cavity field is initially in a coherent state $|\alpha\rangle$ thus the total system is in $|\psi_1\rangle\equiv \frac{1}{\sqrt{1+|\varepsilon_{0}|^2}}(|g\rangle|\alpha\rangle+\varepsilon_0|e\rangle|\alpha\rangle)$. After an interaction time $t =\frac{\pi\Delta}{2g^2} $, the atom-cavity system evolves to
\be
|\psi_1\rangle\rightarrow \frac{1}{\sqrt{1+|\varepsilon_{0}|^2}}(|g\rangle|i\alpha\rangle+\varepsilon_{0}|e\rangle|-i\alpha\rangle).
\ee
Again the atomic transition is resonant with the classical field but far away from the cavity field. After a given time and performing a measurement on the atom we have
\be\label{7}
|\psi_1'\rangle=\frac{1}{\sqrt{M^{(2)}}}(A_0^{1}|-i\alpha\rangle+A_1^{1}|i\alpha\rangle),
\ee
where $A_0^1=-\varepsilon_0\varepsilon_1^*$, $A_1^1=1$ and $M^{(2)}=(A_{0}^{1})^{2}+(A_{1}^{1})^{2}+2Re(A_{0}^{1}A_{1}^{1})e^{-2|\alpha|^2}$. Further the phase shifter $\hat{\mathcal{P}}=e^{-i\frac{\pi}{2}\hat{a}^{\dag}\hat{a}}$ transforms the state Eq.(\ref{7}) as
\be\label{9}
\hat{\mathcal{P}}|\psi_1'\rangle=\frac{1}{\sqrt{M^{(2)}}}(A_0^{1}|\alpha\rangle+A_1^{1}|-\alpha\rangle).
\ee
Similar to the first step $N=1$, the atom interacts alternately with a (resonant) classical field and with the (dispersive) cavity field $N=2,3,...$. The cavity field allows an arbitrary displacement operation during the process, after the measuring on the atom which find the atom in the ground state $|g\rangle$, we can generate superpositions of coherent states with adjustable weighting factors. Whence in step $N=2$, we have
\be\label{3}
|\psi'_2\rangle=\frac{1}{\sqrt{M^{(3)}}}(A_0^2|2\alpha\rangle+A_1^2|0\rangle+A_2^2|-2\alpha\rangle),
\ee
in which $A_0^2=1$, $A_1^2=-(\varepsilon_0\varepsilon_2^*+\varepsilon_0\varepsilon_1^*)$, $A_2^2=-\varepsilon_1\varepsilon_2^*$ and the normalization factor is
\be
M^{(3)}=(A_{0}^{2})^{2}+(A_{1}^{2})^{2}+(A_{2}^{2})^{2}+2Re(A_{0}^{2}A_{1}^{2})e^{-2|\alpha|^2}+2Re(A_{1}^{2}A_{2}^{2})e^{-2|\alpha|^2}+2Re(A_{0}^{2}A_{2}^{2})e^{-8|\alpha|^2}.
\ee
Acting displacement operator $\hat{D}(\beta)=e^{\beta \hat{a}^{\dag}-\beta^*\hat{a}}$ with property $\hat{D}(\beta)|\alpha\rangle=|\alpha+\beta\rangle$, onto state Eq.(\ref{3}) leads to the following superposition of coherent states
\be\label{8}
\hat{D}(\beta)|\psi'_2\rangle=\frac{1}{{\sqrt{M^{(3)}}}}(A_0^2|2\alpha+\beta\rangle+A_1^2|\beta\rangle+A_2^2|-2\alpha+\beta\rangle).
\ee
For $N=3$ the phase shifter $\hat{\mathcal{P}}=e^{-i\frac{\pi}{2}\hat{a}^{\dag}\hat{a}}$ acting on $|\psi_3\rangle$, yields the final state
\be\label{10}
|\psi'_3\rangle=\frac{1}{\sqrt{M^{(4)}}}(A_0^3|3\alpha\rangle+A_1^3|\alpha\rangle+A_2^3|-\alpha\rangle+A_3^3|-3\alpha\rangle),
\ee
where the normalization factor is
\be
\begin{array}{l}
M^{(4)}=(A_{0}^{3})^{2}+(A_{1}^{3})^{2}+(A_{2}^{3})^{2}+(A_{3}^{3})^{2}+2A_{0}^{3}A_{1}^{3}e^{-2|\alpha|^2}
+2A_{0}^{3}A_{2}^{3}e^{-8|\alpha|^2}+2A_{0}^{3}A_{3}^{3}e^{-18|\alpha|^2}\\
~~~~~~~+2A_{1}^{3}A_{2}^{3}e^{-2|\alpha|^2}+2A_{1}^{3}A_{3}^{3}e^{-8|\alpha|^2}+2A_{2}^{3}A_{3}^{3}e^{-2|\alpha|^2},
\end{array}
\ee
and $A_{0}^{3}$, $A_{1}^{3}$, $A_{2}^{3}$ and $A_{3}^{3}$ are functions of $\varepsilon_k$ ({k=0,...,3}). Note that we used the superscripts (2), (3) and (4) for qubit, qutrit and qufit like states respectively.
\subsection{Generation of ECS's Using Beam Splitter}
\begin{figure}[ht]
\centerline{\includegraphics[width=8cm]{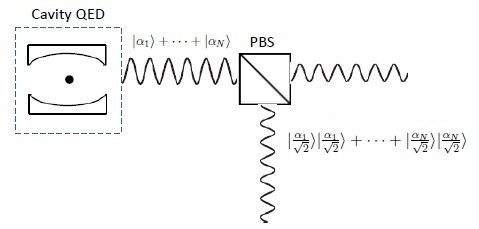}}
\caption{\small Experimental set up for generating ECS \label{set}}
\end{figure}

Here we use $50-50$ beam splitter and generate two-mode ECS's. The polarizing beam splitter (PBS) is commonly made by cementing together two birefringent materials like calcite or quartz. It has the property of splitting a light beam into its orthogonal linear polarizations.
The beam splitter interaction is given by the unitary transformation as
\be
\hat{B}_{i-1,i}(\theta)=\exp[\theta(\hat{a}_{i-1}^{\dag}\hat{a}_{i}-\hat{a}_{i}^{\dag}\hat{a}_{i-1})],
\ee
which $\hat{a}_{i-1}$, $\hat{a}_{i}$, $\hat{a}^{\dag}_{i-1}$ and $\hat{a}^{\dag}_{i}$ are the annihilation and creation operators of the field mode $i-1$ and $i$, respectively. Using Baker-Hausdorf formula, the action of the  $50-50$ beam splitter i.e. $\theta=\pi/4$ on two modes $1$ and $2$, can be expressed as
\be
\hat{B}_{1,2}(\pi/4)|\alpha\rangle_{_{1}}|0\rangle_{_{2}}=|\frac{\alpha}{\sqrt{2}}\rangle_{_{1}}|\frac{\alpha}{\sqrt{2}}\rangle_{_{2}}.
\ee
This  result says that like classical light wave where the incident intensity is evenly divided between the two output beams, e.g. half the incident average photon number, $\frac{|\alpha|^{2}}{2}$, emerges in each beam. Note that the output is not entangled. For producing ECS suppose that our input state be a superposition of different coherent states as Eqs.(\ref{9}), (\ref{8}) and (\ref{10}). Following the procedure above, we may then, obtain the output states as (see figure \ref{set})
\be\label{qubit}
\begin{array}{l}
|\Psi^{(2)}\rangle
=\frac{1}{\sqrt{M^{(2)}}}(A_0^1|\frac{\alpha}{\sqrt{2}}\rangle_{_{1}}|\frac{\alpha}{\sqrt{2}}\rangle_{_{2}}+
A_1^1|\frac{-\alpha}{\sqrt{2}}\rangle_{_{1}}|\frac{-\alpha}{\sqrt{2}}\rangle_{_{2}}),\\
|\Psi^{(3)}\rangle=
\frac{1}{\sqrt{M^{(3)}}}(A_0^2|\frac{2\alpha+\beta}{\sqrt{2}}\rangle_{_{1}}|\frac{2\alpha+\beta}{\sqrt{2}}\rangle_{_{2}}+
A_1^2|\frac{\beta}{\sqrt{2}}\rangle_{_{1}}|\frac{\beta}{\sqrt{2}}\rangle_{_{2}}+A_2^2|\frac{-2\alpha+\beta}{\sqrt{2}}\rangle_{_{1}}|\frac{-2\alpha+\beta}{\sqrt{2}}\rangle_{_{2}}),
\end{array}
\ee
and
\be\label{qufit}
\begin{array}{l}
|\Psi^{(4)}\rangle=
\frac{1}{\sqrt{M^{(4)}}}(A_0^3|\frac{3\alpha}{\sqrt{2}}\rangle_{_{1}}|\frac{3\alpha}{\sqrt{2}}\rangle_{_{2}}+
A_1^3|\frac{\alpha}{\sqrt{2}}\rangle_{_{1}}|\frac{\alpha}{\sqrt{2}}\rangle_{_{2}}+A_2^3|\frac{-\alpha}{\sqrt{2}}\rangle_{_{1}}|\frac{-\alpha}{\sqrt{2}}\rangle_{_{2}}+A_3^3|\frac{-3\alpha}{\sqrt{2}}\rangle_{_{1}}|\frac{-3\alpha}{\sqrt{2}}\rangle_{_{2}}).
\end{array}
\ee
The states $|\Psi^{(2)}\rangle$, $|\Psi^{(3)}\rangle$ and $|\Psi^{(4)}\rangle$ are in general entangled states. We are interested in the amount of bipartite entanglement between the two modes in the
above states. To do this, we use concurrence measure \cite{Wootters1,Wootters2,Akhtarshenas}. Let us consider the general form of  bipartite quantum state in the usual orthonormal basis $|e_{i}\rangle$ as
\be\label{state1}
|\psi\rangle=\sum_{i=1}^{d_{1}}\sum_{j=1}^{d_{2}}a_{ij}|e_{i}\otimes e_{j}\rangle,
\ee
where $d_{1}$ and $d_{2}$  are  dimensions of first and second part respectively. The norm of concurrence vector is defined as
\be
C=\sqrt{\sum_{a=1}^{d_{1}(d_{1}-1)/2}~~\sum_{b=1}^{d_{2}(d_{2}-1)/2}|C_{ab}|^{2}},
\ee
where $C_{ab}=\langle\psi|\widetilde{\psi}_{ab}\rangle$, $|\widetilde{\psi}_{ab}\rangle=(L_{a}\otimes L_{b})|\psi^{*}\rangle$, and $L_{a}$ and $L_{b}$ are the generators of $SO(d_{1})$ and $SO(d_{2})$ respectively. Note that $|\psi^{*}\rangle$ is complex conjugate of  $|\psi\rangle$.
The concurrence in terms of coefficients $a_{ij}$ is
\be\label{cc}
C = 2  \sqrt {\sum\limits_{i < j}^{d_1 } {\sum\limits_{k < l}^{d_2 } {\left| {a_{ik} a_{jl}  - a_{il} a_{jk} } \right|^2 } } }.
\ee
{\bf{Qubit case:}}
Two non-orthogonal coherent states $|\frac{\alpha}{\sqrt{2}}\rangle$ and
$|\frac{-\alpha}{\sqrt{2}}\rangle$ are assumed to be linearly independent and span a two-dimensional subspace of the Hilbert space. Returning to the particular problem at hand, one can transform the  $|\Psi^{(2)}\rangle$  to a form analogous to Eq. (\ref{state1}) by defining the orthonormal basis
\be\label{base}
\begin{array}{l}
|0\rangle=|\frac{\alpha}{\sqrt{2}}\rangle,\\
|1\rangle=\frac{1}{\sqrt{1-p^{2}}}(|\frac{-\alpha}{\sqrt{2}}\rangle-p|\frac{\alpha}{\sqrt{2}}\rangle),
\end{array}
\ee
in which $p=\langle\frac{\alpha}{\sqrt{2}}|\frac{-\alpha}{\sqrt{2}}\rangle$. Clearly the first state in Eq.(\ref{qubit}) is a qubit like state. Using Eq.(\ref{cc}) concurrence can be obtained as
\be
C^{(2)}=\frac{2|\varepsilon_0\varepsilon_1^*|(1-p^2)}{1+|\varepsilon_0\varepsilon_1^*|^2-2Re(\varepsilon_0\varepsilon_1^*)p^2}.
\ee
By assuming $\varepsilon_0\varepsilon_1^*=1$ the concurrence is $C^{(2)}=1$, meaning that if $\varepsilon_0\varepsilon_1^*=1$, independent of values of $p$, concurrence is maximal i.e. the state $|\Psi^{(2)}\rangle$ is in the category of Bell states, $|\Psi^{(2)}\rangle=\frac{1}{\sqrt{2}}(|00\rangle-|11\rangle)$\cite{najarbashi1}.\\
{\bf{Qutrit case:}}
Three non-orthogonal coherent states $|\frac{\beta}{\sqrt{2}}\rangle$,
$|\frac{2\alpha+\beta}{\sqrt{2}}\rangle$ and $|\frac{-2\alpha+\beta}{\sqrt{2}}\rangle$ are assumed to be linearly independent and span a three-dimensional subspace of the Hilbert space. Similarly for qutrit case the orthonormal basis are defined as
\be\label{base}
\begin{array}{l}
|0\rangle=|\frac{\beta}{\sqrt{2}}\rangle,\\
|1\rangle=\frac{1}{\sqrt{1-p_{1}^{2}}}(|\frac{2\alpha+\beta}{\sqrt{2}}\rangle-p_1|\frac{\beta}{\sqrt{2}}),\\
|2\rangle=\sqrt{\frac{1-p_{1}^{2}}{1-p_{1}^{2}-p_{2}^{2}-p_{3}^{2}+2p_1p_2p_3}}(|\frac{-2\alpha+\beta}{\sqrt{2}}\rangle+(\frac{p_1p_3-p_2}{1-p_{1}^{2}})|\frac{2\alpha+\beta}{\sqrt{2}}\rangle+(\frac{p_1p_2-p_3}{1-p_{1}^{2}})|\frac{\beta}{\sqrt{2}}\rangle),
\end{array}
\ee
in which $p_1=\langle\frac{\beta}{\sqrt{2}}|\frac{2\alpha+\beta}{\sqrt{2}}\rangle$, $p_2=\langle\frac{2\alpha+\beta}{\sqrt{2}}|\frac{-2\alpha+\beta}{\sqrt{2}}\rangle$ and $p_3=\langle\frac{\beta}{\sqrt{2}}|\frac{-2\alpha+\beta}{\sqrt{2}}\rangle$.
In \cite{Zeng} the optimal values of $\varepsilon_0$, $\varepsilon_1$, and $\varepsilon_2$ determined to be $-0.8200$,
$2.1184$, and $-0.4720$, respectively. Then by Eq.(\ref{cc}) concurrence reads
\be
\begin{array}{l}
C^{(3)}(\alpha)=
 \frac{2\sqrt{
  4.64461 + 14.098 p^{14} - 17.6206 p^{12}-
   6.05322p^{10}+ 24.965 p^{8}-22.5563 p^6+ 2.59035 p^4-0.0677901p^2}}{
 1.99954p^8 + 7.28968 p^2+ 3.8224},
\end{array}
\ee
we note that $p=e^{-\alpha^2}$. Similarly by definition the orthonormal basis $|0\rangle$, $|1\rangle$, $|2\rangle$ and $|3\rangle$ for qufit like state Eq.(\ref{qufit}) and by assumption $A_0^3=A_2^3=A_3^3=-1$ and $A_1^3=1$ the concurrence can be easily obtained as a function of $\alpha$.
\begin{figure}[ht]
\centerline{\includegraphics[width=9cm]{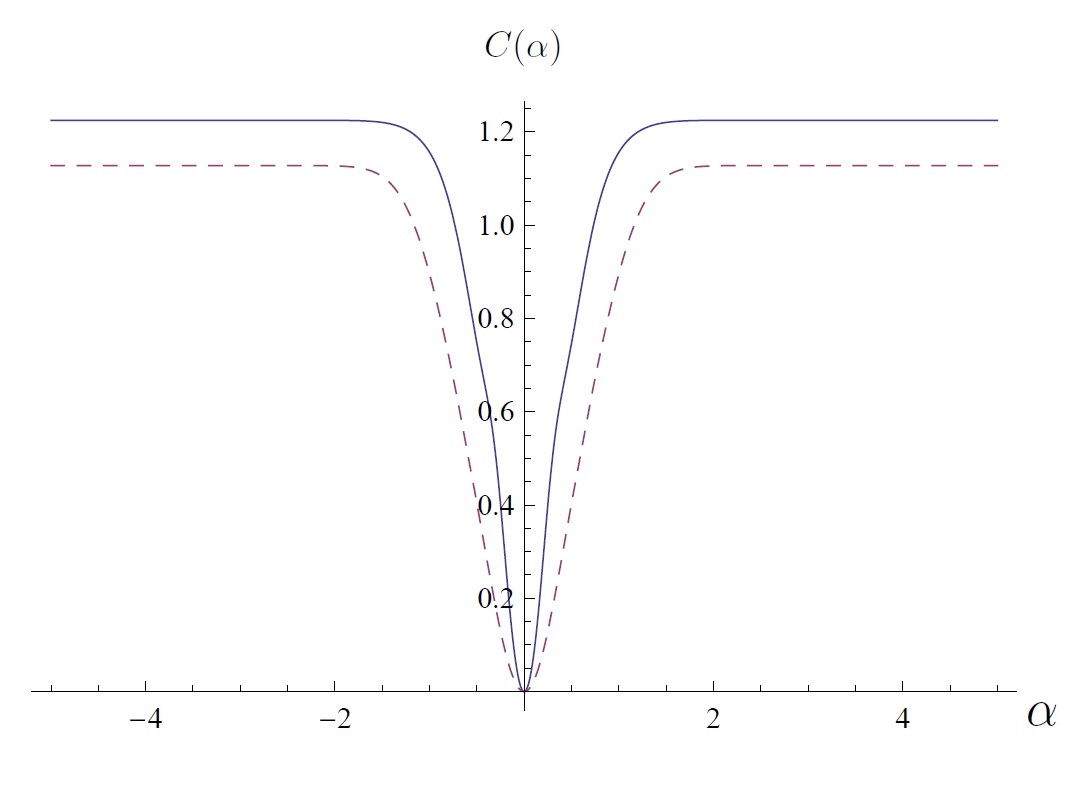}}
\caption{\small {Concurrences $C^{(3)}(\alpha)$ (dashed line) and $C^{(4)}(\alpha)$ (full line) as a function of $\alpha$} \label{concurrence} }
\end{figure}
Figure \ref{concurrence} indicates the behaviour of concurrences $C^{(3)}(\alpha)$ and $C^{(4)}(\alpha)$ as a function of $\alpha$, where for simplicity we assumed that $\alpha$ is real. Figure \ref{concurrence} shows that except for $\alpha\rightarrow 0$, the states $\Psi^{(3)}$ and $\Psi^{(4)}$ are entangled. Moreover, for large $\alpha$, concurrence reaches its maximum value i.e. $C^{(3)}(\alpha)=1.128$ for qutrit like state and $C^{(4)}(\alpha)=1.224$ for qufit like state.
\section{Effects of Noise on ECS's}
\begin{figure}[ht]
\centerline{\includegraphics[width=9cm]{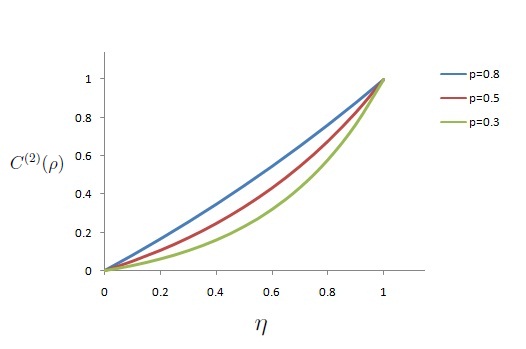}}
\caption{\small {(Color online) Concurrence $C^{(2)}(\rho)$ as a function of $\eta$ for given $p=0.3$, $p=0.5$ and $p=0.8$ } \label{MIXED} }
\end{figure}

\begin{figure}[ht]
\centerline{\includegraphics[width=9cm]{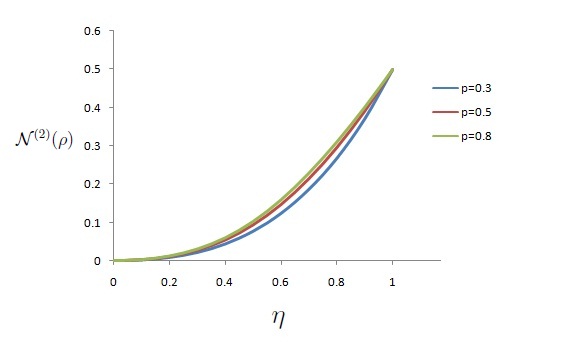}}
\caption{\small {(Color online) Negativity of two qubit like state $\mathcal{N}^{(2)}(\rho)$ as a function of $\eta$ for given $p=0.3$, $p=0.5$ and $p=0.8$ } \label{qubitne} }
\end{figure}

\begin{figure}[ht]
\centerline{\includegraphics[width=9cm]{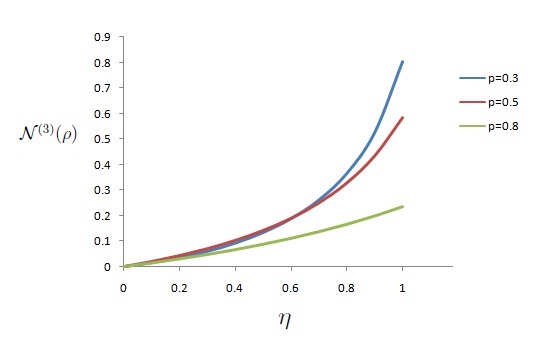}}
\caption{\small {(Color online) Negativity of two qutrit like state $\mathcal{N}^{(3)}(\rho)$ as a function of $\eta$ for given $p=0.3$, $p=0.5$ and $p=0.8$ } \label{Negativity} }
\end{figure}

\begin{figure}[ht]
\centerline{\includegraphics[width=9cm]{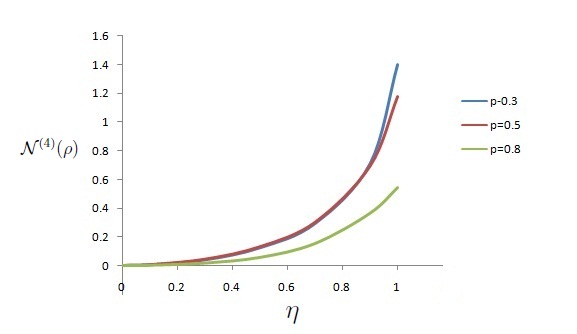}}
\caption{\small {(Color online) Negativity of two qufit like state $\mathcal{N}^{(4)}(\rho)$ as a function of $\eta$ for given $p=0.3$, $p=0.5$ and $p=0.8$ } \label{qufne} }
\end{figure}

Let us assume that the mode $1$ travel through a noisy channel characterized by
\be
|\alpha\rangle_{1}|0\rangle_E\rightarrow |\sqrt{\eta}\alpha\rangle_{1}|\sqrt{1-\eta}\rangle_E,
\ee
where the second state now refers to the environment and $\eta$ is the noise parameter, which gives the fraction of photons that survives the noisy channel \cite{Yao}. Here the effects of noise on entangled qubit, qutrit and qufit like states are investigated.\\
{\bf{Qubit case:}}
Starting from a state $|\Psi^{(2)}\rangle$ the state after traveling through the noisy channel becomes
\be
\begin{array}{l}
|\Psi'^{(2)}\rangle=\frac{1}{\sqrt{M^{(2)}}}(A_0^1|\sqrt{\eta}\beta,\sqrt{\eta}\beta\rangle_{_{1,2}}|\sqrt{1-\eta}\beta,\sqrt{1-\eta}\beta\rangle_{_E}\\
~~~~~~~~~~+A_1^1|-\sqrt{\eta}\beta,-\sqrt{\eta}\beta\rangle_{_{1,2}}|-\sqrt{1-\eta}\beta,-\sqrt{1-\eta}\beta\rangle_{_E}),\\
\end{array}
\ee
in which $\beta=\frac{\alpha}{\sqrt{2}}$. In order to study the noise effect on entanglement between modes $1$ and $2$, we should trace out the environment mode $E$ by partial trace, i.e. $\rho_{12}=Tr_{E}(|\Psi'^{(2)}\rangle\langle\Psi'^{(2)}|)$, then reduced density matrix in orthogonal basis $|0\rangle$ and $|1\rangle$ reads
\be
\rho_{12}^{(2)}=\frac{1}{2-2p^2}\left(
  \begin{array}{cccc}
    a_{11} & a_{12} & a_{12} & a_{14} \\
    a_{12} & a_{22} & a_{22} & a_{24} \\
    a_{12} & a_{22} & a_{22} & a_{24} \\
    a_{14} & a_{24} & a_{24} & a_{44} \\
  \end{array}
\right),
\ee
where
\be\begin{array}{l}
a_{11}=1-2p^2+p^{4\eta},\\
a_{12}=p^{-\eta} \sqrt{1-p^{2\eta}}(-p^2 + p^{4 \eta}),\\
a_{14}=p^2-p^{2-2\eta} + p^{2\eta}-p^{4\eta},\\
a_{22}=p^{2\eta}-p^{4\eta},\\
a_{24}=p^{\eta} (1-p^{2\eta})^{3/2},\\
a_{44}=(1-p^{2\eta})^2.
\end{array}
\ee
Clearly this state is a two qubit mixed state and one of the suitable measure to evaluate the amount of entanglement is concurrence.
For any two-qubit mixed state, concurrence is defined as $C=\max\{0,\lambda_{1}-\lambda_{2}-\lambda_{3}-\lambda_{4}\}$ where the $\lambda_{i}$'s are the non-negative eigenvalues, in decreasing order, of the Hermitian matrix $R=\sqrt{\sqrt{\rho}\tilde{\rho}\sqrt{\rho}}$,
with $\tilde{\rho}=(\sigma_{y}\otimes\sigma_{y})\rho^{*}(\sigma_{y}\otimes\sigma_{y})$
in which $\rho^{*}$ is the complex conjugate of $\rho$ when it is expressed in a standard basis and $\sigma_{y}$ represents the usual second Pauli matrix in a local basis $\{|0\rangle, |1\rangle\}$ \cite{Wootters1}.\\
Figure \ref{MIXED} shows the behaviour of concurrence for $\rho_{12}^{(2)}$ as a function of $\eta$.
 From figure \ref{MIXED}, we see that concurrence is decreased by decreasing noise parameter $\eta$, it means that entanglement of ECS $\rho_{12}^{(2)}$, is decreased after traveling noisy channel. Moreover by increasing the amplitude of coherent state $\alpha$ (or equivalently decreasing $p$) the concurrence also decreases. In order to check this result and for the latter use, let us calculate the negativity defined as \cite{Werner},
\be
\mathcal{N}(\rho)=\frac{||\rho^{T_B}||_1-1}{2}=|\sum_i \mu_i|,
\ee
where $\rho^{T_B}$ denote the partial transpose and $||\rho^{T_B}||_1= Tr\sqrt{(\rho^{T_B})^{\dag} \rho^{T_B}}$ is the usual trace norm. The second equality comes from the condition $Tr(\rho)=1$ and $\mu_i$'s are the negative eigenvalues of $\rho^{T_B}$.
The behaviour of negativity as a function of noise parameter $\eta$ is shown in figure \ref{qubitne}.
We deduce that the negativity shows the same results as concurrence.\\
{\bf{Qutrit case:}}
For a qutrit like state $|\Psi^{(3)}\rangle$, after traveling through the noisy channel the state becomes
\be\label{5}
\begin{array}{l}
|\Psi'^{(3)}\rangle=\frac{1}{\sqrt{3.8225+5.4e^{-2\alpha^2}+2e^{-8\alpha^2}}}(|\sqrt{\eta}\omega,\sqrt{\eta}\omega\rangle_{1,2}|\sqrt{1-\eta}\omega,\sqrt{1-\eta}\omega\rangle_E\\
~~~~~~~~~~+1.35|\sqrt{\eta}\delta,\sqrt{\eta}\delta\rangle_{1,2}|\sqrt{1-\eta}\delta,\sqrt{1-\eta}\delta\rangle_E\\
~~~~~~~~~~+|\sqrt{\eta}\gamma,\sqrt{\eta}\gamma\rangle_{1,2}|\sqrt{1-\eta}\gamma,\sqrt{1-\eta}\gamma\rangle_E),
\end{array}
\ee
where $|\frac{\beta}{\sqrt{2}}\rangle\equiv|\delta\rangle$, $|\frac{2\alpha+\beta}{\sqrt{2}}\rangle\equiv|\omega\rangle$ and $|\frac{-2\alpha+\beta}{\sqrt{2}}\rangle\equiv|\gamma\rangle$.
Now we can investigate noise effect on entanglement of two-mode ECS hence we must calculate entanglement among modes 1 and 2. If we trace out the environment mode $E$, (i.e. $\rho_{12}^{(3)}=Tr_{E}(|\Psi'^{(3)}\rangle\langle\Psi'^{(3)}|)$), the reduced density matrix reads
\be\label{6}
\begin{array}{l}
\rho_{1,2}^{(3)}=
\frac{1}{3.8225+5.4p^2+2p^8}(|\sqrt{\eta}\omega,\sqrt{\eta}\omega\rangle\langle\sqrt{\eta}\omega,\sqrt{\eta}\omega|\\
~~~~~+1.35A(|\sqrt{\eta}\omega,\sqrt{\eta}\omega\rangle\langle\sqrt{\eta}\delta,\sqrt{\eta}\delta|+|\sqrt{\eta}\delta,\sqrt{\eta}\delta\rangle\langle\sqrt{\eta}\omega,\sqrt{\eta}\omega|)\\
~~~~~+1.35B(|\sqrt{\eta}\delta,\sqrt{\eta}\delta\rangle\langle\sqrt{\eta}\gamma,\sqrt{\eta}\gamma|+|\sqrt{\eta}\gamma,\sqrt{\eta}\gamma\rangle\langle\sqrt{\eta}\delta,\sqrt{\eta}\delta|)\\
~~~~~+D(|\sqrt{\eta}\omega,\sqrt{\eta}\omega\rangle\langle\sqrt{\eta}\gamma,\sqrt{\eta}\gamma|+|\sqrt{\eta}\gamma,\sqrt{\eta}\gamma\rangle\langle\sqrt{\eta}\omega,\sqrt{\eta}\omega|)\\
~~~~~+1.8225|\sqrt{\eta}\delta,\sqrt{\eta}\delta\rangle\langle\sqrt{\eta}\delta,\sqrt{\eta}\delta|+|\sqrt{\eta}\gamma,\sqrt{\eta}\gamma\rangle\langle\sqrt{\eta}\gamma,\sqrt{\eta}\gamma|,
\end{array}
\ee
where $A$, $B$ and $D$ are defined as
\be
\begin{array}{l}
A=\langle\sqrt{1-\eta}\delta,\sqrt{1-\eta}\delta|\sqrt{1-\eta}\omega,\sqrt{1-\eta}\omega\rangle=p^{2(1-\eta)},\\ B=\langle\sqrt{1-\eta}\gamma,\sqrt{1-\eta}\gamma|\sqrt{1-\eta}\delta,\sqrt{1-\eta}\delta\rangle=p^{2(1-\eta)},\\ D=\langle\sqrt{1-\eta}\gamma,\sqrt{1-\eta}\gamma|\sqrt{1-\eta}\omega,\sqrt{1-\eta}\omega\rangle=p^{8(1-\eta)},
\end{array}
\ee
and $p=e^{-\alpha^2}$. The negativity as a function of $\eta$ is plotted in figure \ref{Negativity}.
Similarly the behaviour of negativity of decohered two qufit like state is shown in figure \ref{qufne}.
Figures \ref{Negativity} and \ref{qufne} show us that negativity is decreased by decreasing noise parameter $\eta$, i.e. entanglement of ECS is decreased after traveling noisy channel. Moreover it can be seen that by increasing the amplitude of coherent state $\alpha$ ($|\alpha|\rightarrow\infty$ or equivalently $p\rightarrow 0$) ECS's decohere faster i.e. the entanglement between modes $1$ and $2$ will rapidly disappear.
\section{Noise Effects and Monogamy Inequality}
\begin{figure}[ht]
\centerline{\includegraphics[width=9cm]{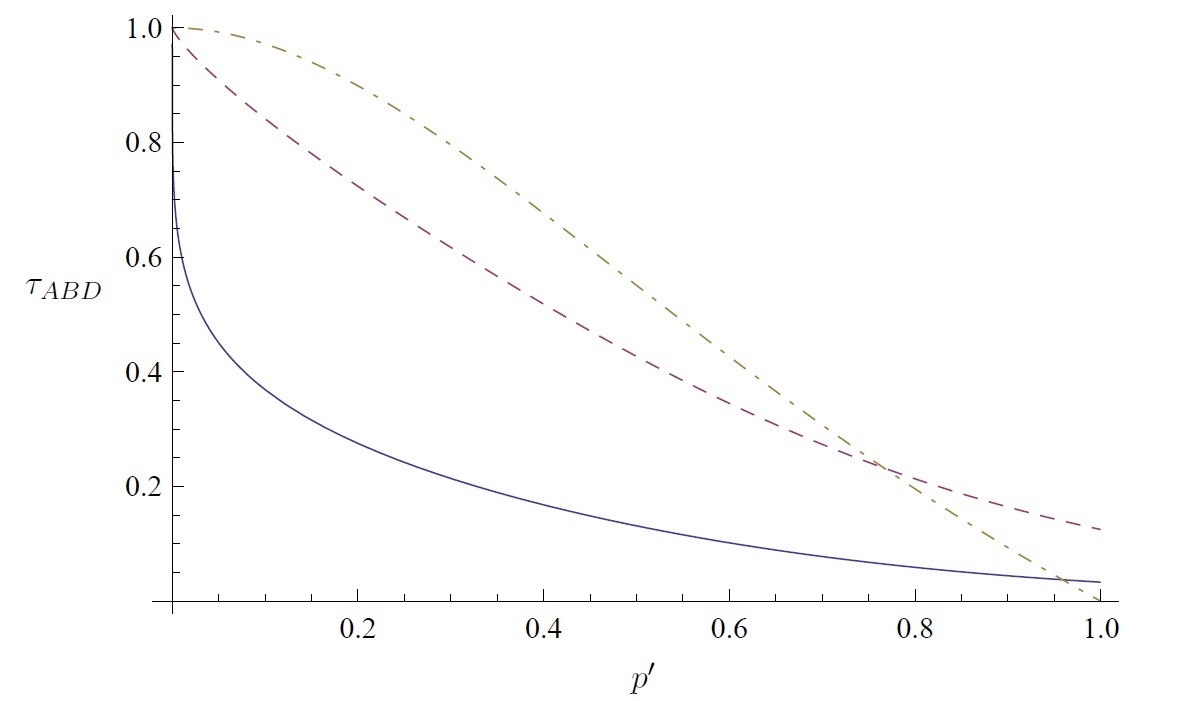}}
\caption{\small {$\tau_{ABD}=C_{A(BD)}^2-C_{AB}^2-C_{AD}^2$ as a function of $p'$ for $\eta=0.1$ (full line), $\eta=0.4$ (dashed line) and $\eta=1$ (dot-dashed line).} \label{t} }
\end{figure}

Monogamy is the another property of quantum entanglement which is expressed by inequality: \cite{coffman,Ou,kim1,sanders1}.
\be
\label{monogamy}
C_{A(BD)}^{2}\geq C_{AB}^{2}+C_{AD}^{2},
\ee
where $C_{AB}$ and $C_{AD}$ are the concurrences of the reduced density matrices of $\rho_{AB}$ and $\rho_{AD}$ respectively and $C_{A(BD)}$ is the concurrence of pure state $|\psi\rangle_{ABD}$ with respect to two partitions  $A$ and $BD$. Monogamy inequality is hold for any multi qubit states but there are some examples in
qutrit quantum systems violating concurrence-based monogamy inequality. We see here how the difference of the two sides of monogamy inequality ($\tau_{ABD}=C_{A(BD)}^2-C_{AB}^2-C_{AD}^2$) vary as a function of noise parameter $\eta$. Let us consider the state Eq.(\ref{9}), after transforming two beam splitters \cite{najarbashi1} it reads
\be
\begin{array}{l}\label{11}
\hat{B}_{23}(\frac{\pi}{4})\hat{B}_{12}(\cos^{-1}(\frac{1}{\sqrt{3}}))(\frac{1}{\sqrt{M^{(2)}}}(|-\alpha\rangle_1-|\alpha\rangle_1)\otimes|0\rangle_2\otimes|0\rangle_3)\\
~~~~~~~~~~~~~~~=\frac{1}{\sqrt{M^{(2)}}}(|\frac{-\alpha}{\sqrt{3}}\rangle_1|\frac{-\alpha}{\sqrt{3}}\rangle_2|\frac{-\alpha}{\sqrt{3}}\rangle_3-|\frac{\alpha}{\sqrt{3}}\rangle_1|\frac{\alpha}{\sqrt{3}}\rangle_2|\frac{\alpha}{\sqrt{3}}\rangle_3),
\end{array}
\ee
in which $M^{(2)}=2-2p'^3$ and $p'=\langle\frac{\alpha}{\sqrt{3}}|\frac{-\alpha}{\sqrt{3}}\rangle$. Using concurrence formula for two qubit mixed state which is mentioned in section 3, we have \be
\begin{array}{l}
C_{AB}=C_{AD}=\frac{p'(1+p')}{1+p'+p'^2},
\end{array}
\ee
and from Eq.(\ref{cc}), it can be found that
\be
C_{A(BD)}=\frac{\sqrt{(1-p'^2)(1-p'^4)}}{1-p'^3}.
\ee
We analyze in this manner the effects of noise on parameter $\tau_{ABD}$. The state (\ref{11}) after transmitting through a noisy channel is
\be
\begin{array}{l}
|\varphi\rangle=\frac{1}{\sqrt{M^{(2)}}}(|-\sqrt{\eta}\beta,-\sqrt{\eta}\beta,-\sqrt{\eta}\beta\rangle|-\sqrt{1-\eta}\beta,-\sqrt{1-\eta}\beta,-\sqrt{1-\eta}\beta\rangle_E\\
~~~~~~~~~~~~~+|\sqrt{\eta}\beta,\sqrt{\eta}\beta,\sqrt{\eta}\beta\rangle|\sqrt{1-\eta}\beta,\sqrt{1-\eta}\beta,\sqrt{1-\eta}\beta\rangle_E),
\end{array}
\ee
then we have
\be
\begin{array}{l}
C_{AB}=C_{AD}=\frac{p'^{3+\eta}(1-p'^{2\eta})}{p'^{3\eta}-p'^6},\\
C_{A(BD)}=\frac{\sqrt{(1-p'^{2\eta})(1-p'^{2(3-\eta)})}}{1-p'^{6-3\eta}}.
\end{array}
\ee
The behaviour of $\tau_{ABD}=C_{A(BD)}^2-C_{AB}^2-C_{AD}^2$ as a function of $\eta$ is represented in figure \ref{t}.
Positivity of $\tau_{ABD}$ shows that monogamy inequality is hold for qubit states. Figure \ref{t}, shows that if $\eta=1$ (lossless channel) and $p'<0.76$  the maximum violation occurs. Moreover, by decreasing $\eta$, the amount of $\tau_{ABD}$ is reduced for $p'<0.76$, but there are $p'$ for which the crossing occurs, i.e., $\tau_{ABD}$ may be  decreased by increasing $\eta$.
\\
As an example which violates the monogamy inequality, consider the following  imbalanced qutrit like ECS
\be\label{qutrit}
|\psi\rangle=\frac{1}{\sqrt{M'^{(3)}}}(\mu_1(|\alpha\beta\gamma\rangle+|\beta\gamma\alpha\rangle+|\gamma\alpha\beta\rangle)+
\mu_2(|\alpha\gamma\beta\rangle+|\beta\alpha\gamma\rangle+|\gamma\beta\alpha\rangle)),
\ee
where $M'^{(3)}$ is the normalization factor. We note that three first terms are considered as even permutation and the other are considered as odd permutation. From  Eq.(\ref{base}) if we assume $\mu_1=-\mu_2=1$ and $p_1,p_2$ and $p_3$ tend to zero ($p_1,p_2,p_3\rightarrow 0$), it means that $|\alpha\rangle$, $|\beta\rangle$ and $|\gamma\rangle$ are orthogonal and we have
\be
|\psi\rangle=\frac{1}{\sqrt{6}}(|012\rangle-|021\rangle+|201\rangle-|210\rangle+|120\rangle-|102\rangle).
\ee
It has been shown that in \cite{Ou,sanders1}
\be
C_{AB}^2+C_{AD}^2=2\geq \frac{4}{3}=C_{A(BD)}^2.
\ee
which is obviously a violation of the monogamy inequality Eq.(\ref{monogamy}) for qutrit case.
\section{Conclusion}
In summary, we used the Jaynes-Cummings model to produce two and three-mode ECS's via beam splitters. We also found the amount of entanglement for generated qubit, qutrit and qufit like ECS's by concurrence and negativity. Our quantitative calculations of entanglement showed that for qubit like ECS by assuming $\varepsilon_0\varepsilon_1^*=1$ the concurrence independent of values of $p$ is maximal ($C^{(2)}=1$) that is the state $|\Psi^{(2)}\rangle$ is in the category of Bell states. For qutrit and qufit like ECS's the concurrence reachs its maximum for large $\alpha$, i.e. $C^{(3)}(\alpha)=1.128$ and $C^{(4)}(\alpha)=1.224$. Moreover, we particulary focused on decoherence properties of ECS's due to noisy channels. To this aim we used negativity. It was shown that, negativity is decreased by decreasing noise parameter $\eta$, i.e. entanglement of ECS's are decreased after traveling through noisy channel. Finally, it was analyzed that if $\eta=1$ (lossless channel) the maximum violation of monogamy inequality holds and $\tau_{ABD}$ is reduced by decreasing noise parameter $\eta$ for $p'<0.76$. We introduced a qutrit like ECS violating the monogamy inequality.\\

\textbf{Acknowledgments}\\
The authors also acknowledge the support from the University of Mohaghegh Ardabili.

\end{document}